\documentstyle[12pt,psfig]{article}

\newcommand{\tr}{\mbox{tr}}

\begin{document}
\begin{flushright}
SLAC-PUB-8031\\
December 1998
\end{flushright}
\begin{center}
{\LARGE 
{Strong $WW$ interactions at the LHC}
\footnote{Research partially supported
by the Department of Energy under contract DE-AC03-76SF00515
and the Spanish CICYT under contract AEN97-1693.
}}\\[2ex]

Antonio Dobado\\
\hspace*{.2in}
{\small\em Departamento de F\'{\i}sica Te\'orica} \\
{\small\em Universidad Complutense de Madrid.
28040 Madrid, Spain} 

\vspace*{.2in}
J. R. Pel\'aez\footnote{E-mail:pelaez@slac.stanford.edu. On
leave of absence from the Departamento de F\'{\i}sica Te\'orica.}\\
\hspace*{.2in}
{\small\em Stanford Linear Accelerator Center}\\
{\small\em Stanford University,
Stanford, California 94309. U.S.A.}

\vspace{2cm}
Invited Talk to the XXVI International Meeting on Fundamental Physics\\
Illa da Toxa, Galicia, SPAIN. 1-5 June 1998

\end{center}


\abstract{We present a brief pedagogical 
introduction to the
Effective Electroweak Chiral Lagrangians, which
provide a model independent description of the WW interactions 
in the strong regime. When it is
complemented with some unitarization
or a dispersive approach, this formalism
allows the study of the general strong scenario expected at the LHC,
 including resonances. }

\section{Introduction}

As  is well known, the Standard Model (SM),
which is a $SU(3)_C\times SU(2)_L \times U(1)_Y$ quantum gauge
theory, is able to describe all our knowledge
about the strong and electroweak 
interactions, even at the high level 
of precision reached at LEP (see for instance \cite{Mana}).  
The SM can be divided in three sectors: The 
first one is the matter content (quarks and leptons), whose 
elementary particles interact
 between themselves by mediating bosons
that belong to the second sector. 
These fields  are the eight gluons associated to the $SU(3)_C$ 
group of the strong interactions as well as the $W^+, W^-$ and $Z$
 bosons together with the photon, which are associated
to the $SU(2)_L \times U(1)_Y$ group of the electroweak 
interactions. Finally, there is the so-called
 Symmetry Breaking Sector (SBS). It is responsible for the spontaneous
 symmetry breaking of the electroweak gauge group $SU(2)_L \times U(1)$ 
down to the electromagnetic group $U(1)_{em}$. 
Through the Higgs mechanism it provides masses for the $W^+, W^-$ 
and $Z$ bosons while leaving the photon massless. In addition,
the SBS is also connected with the matter sector through Yukawa 
couplings which give rise to the quark and lepton masses, quark
mixing (Cabbibo-Kowayashi-Maskawa matrix) and 
eventually to CP violation because of the complex phase in this matrix.

Now we arrive to the first important remark of this lecture. In contrast
 with the matter and the gauge sectors, {\it the SBS is very poorly known
 from the experimental point of view}. In fact, several different 
theoretical scenarios have been widely discussed in the literature.
Generically they can be grouped in three kinds, 
the Minimal Standard Model (MSM), the Minimal 
Supersymmetric Standard Model (MSSM)\cite{susy} 
and QCD-like theories \cite{qcd}. 
 Let us briefly review the main 
features and problems of these scenarios.

\subsection{Minimal Standard Model}

It contains the minimum ingredients to explain the present data.
However, it does not shed much light on possible new physics effects
and it does not address several problems, among others:
\begin{itemize}
\item\
The Higgs potential is introduced {\it ad hoc}. It is not a 
gauge interaction 
as the rest of the the known forces in Nature such as the strong, 
the electroweak
 or even gravity. The origin and nature of this Higgs field remains 
a mystery. 
\item\ Keeping the mass of this scalar field at scales close to the
electroweak symmetry breaking (100 GeV to 1 TeV)
requires a very fine tuning, since
radiative corrections tend to make its mass of the order of the next
new physics scale. 
This is known as the
naturalness problem.
\item\
The electron and top masses fall
five orders of magnitude apart. The problem of why the masses
present such a hierarchy is not addressed.
\item\ There are hints in the literature suggesting that
the simple realization of the Higgs sector in the MSM could indeed be
a trivial (non-interacting) quantum field theory.
\end{itemize}
Nowadays the existence of the Higgs is taken for granted by many people
 as it was the case for the top quark. However the Higgs is not the top.
 By this we mean that the SM model 
would have not been  consistent without the top quark 
(it becomes an anomalous gauge theory), whereas the Higgs boson
is not a theoretical need. It is possible to postulate
 different versions of the SM differing in the SBS  which are
theoretically consistent. Indeed, in most of the physical systems
that present an spontaneous symmetry breaking (like chiral 
symmetry breaking
in QCD or, in solid state physics, the 
Cooper pair formation, magnetization, etc..,), 
there is nothing analogous 
to a fundamental Higgs field.

Nevertheless, for its simplicity, this model is very useful to describe the
data, without additional assumptions.

\subsection{Minimal Supersymmetric Standard Model}

In this model, an additional symmetry relating fermions and bosons 
is introduced. As a consequence the Higgs potential is related to the gauge
 couplings and  the scalar particles appear in a natural way. 
The advantage of this new symmetry is that 
for each fermion loop there is a corresponding boson loop with similar
couplings and masses but opposite sign, thus avoiding the 
naturalness problem. However:

\begin{itemize}
\item\
Nature is {\it not} supersymmetric. ``Soft'' breaking terms must be
 added {\ by hand} in order to break spontaneously the 
$SU(2)_L\times U(1)_Y$
 gauge symmetry, without spoiling too much the cancellations needed to 
solve the naturalness problem. Those terms break supersymmetry explicitly.
\item\
The values of the parameters in those soft breaking terms 
 (more than a hundred) are unknown, and they severely limit
the predictive power of these models. The origin of those soft 
breaking terms are the origin of further speculation.
\item\
Probably the most robust (soft breaking parameter independent) prediction 
is that a Higgs  should appear
below around 120 GeV. Thus this particle could have been produced at LEP. So far nothing has been found, but there is still a small room 
for discovery. However, if nothing of this kind is found at the 
next generation of
colliders the low-energy supersymmetric scenarios
would be in serious trouble.
\end{itemize}

\subsection{QCD-like scenario}
These models mimic the spontaneous chiral symmetry breaking
of QCD and are generically known as Technicolor (TC). 
The Higgs does not exist as a fundamental field although some 
other composite fields with different quantum numbers play  a similar role.
However
\begin{itemize}
\item\
There is no completely consistent and 
universally accepted Technicolor model.
\item\
Predictions are very vague due to the strong nature of the interactions.
\item\
The simplest versions, like a direct rescaling of QCD,
are ruled out by the LEP data or by the appearance of undesired
flavor changing neutral currents.
\end{itemize}

Therefore we arrive to the second main remark of this lecture: {\it it
possible that the SBS of the SM has nothing to do with our current
 theoretical expectations}. 

At this point one could ask which are the main experimental constraints on
 the SBS or, in other words, what we really know about this sector. The 
main pieces of our knowledge are the following \cite{cha}:

\begin{enumerate}

\item First of all there must be a 
physical system coupled to the SM displaying
 a spontaneously symmetry breaking pattern 
from a {\it global} $G$ group to 
a subgroup $H$. This symmetry breaking triggers 
the Higgs mechanism that breaks
 the electroweak {\it gauge} group $SU(2)_L \times U(1)_Y$ down to the
electromagnetic group $U(1)_{em}$. Thus we have 
$SU(2)_L \times U(1)_Y \in G$ and $U(1)_{em} \in H$.

\item Since we need three would-be 
Goldstone bosons in order to give masses
 to the $W^+, W^-$ and the $Z$ gauge bosons, we have $dim~G~-~dim~H~=~3$.

\item Experimentally we know that the $\rho$ parameter (which measures the
 relative strength of the charged and neutral weak currents) is very close
 to one - apart from some radiative corrections proportional to 
the hypercharge
 coupling $g'$ squared. Probably the most natural explanation for this
 fact is to assume that the unbroken $H$ group of the SBS contains the 
so called custodial group $SU(2)_{L+R}$ (as it happens in the MSM). 
Any other assumption leads to some fine tuning. 

With these conditions on $G$ and $H$ it 
is very easy to show that the only possible solution is  
$G=SU(2)_L\times SU(2)_R$ and $H=SU(2)_{L+R}$.

\item Finally, from 
 the muon mean life  it is possible to obtain the dimensional parameter 
$v \simeq 250\,\hbox{GeV}$ 
which sets the scale of the SBS dynamics in the SM.

\end{enumerate}

At this point it is reasonable to think whether it is possible to 
build a model independent description of the SBS. As we will see this can 
be done by using the Effective Electroweak 
Chiral Lagrangian (EChL),
 which is based on a similar formalism used in low-energy hadron physics,
 namely, Chiral Perturbation Theory (ChPT) \cite{libro}. As we will see,
this approach is especially useful
 when the SBS is strongly interacting. 

\section{The Electroweak Chiral Lagrangian}

The EChL provides a phenomenological description of the Goldstone boson
 dynamics associated to the symmetry breaking of $SU(2)_L\times
 SU(2)_R$ down to $SU(2)_{L+R}$. As far as we are not
introducing any other field, it has to be realized
nonlinearly. That will limit the applicability of the
approach up to the energies where the other relevant degrees of 
freedom show up. In the case of strong dynamics, we expect these other 
modes to appear at energies much higher than $v\simeq 250\,\hbox{GeV}$ and the
formalism will be very useful. In contrast, for theories with, for instance,
a light Higgs (as in the MSSM), 
there is no applicability region for this formalism, but 
in that case we will have additional information to disentangle 
the SBS physics when measuring these light modes.

Therefore, we will be assuming a strong SBS. For simplicity, let us
then switch off momentarily 
the gauge fields, whose interactions with the SBS are comparatively weak. 
In such case, no other degrees of freedom are present at low energies
except  the Goldstone bosons $\omega^a(x)$, which will be gathered
in the $SU(2)_{L+R}$ matrix
\begin{equation}
U(x)=\exp\left(\frac{i\omega^a(x)\sigma^a}{v}\right),
\end{equation}
where the $\sigma^a$ are the Pauli matrices.

A low energy expansion of the amplitudes
is nothing but a  derivative expansion of the Lagrangian.
Then, the simplest $G$-invariant Lagrangian relevant 
at low energies (with two derivatives) can be written as 
\begin{equation}
{\cal L}_2=\frac{v^2}{4}\tr \,\partial_{\mu}U\partial^{\mu}U^\dagger.
\label{L2}
\end{equation}
From this Lagrangian it is possible to obtain the exact behavior of the
 elastic low-energy scattering amplitude for the Goldstone bosons. 
Indeed, using the SU(2) and crossing symmetries, any 
 amplitude can be obtained from that 
of $\omega^+\omega^-\rightarrow\omega^0\omega^0$, which is given by
\begin{equation}
A(s,t,u)=\frac{s}{v^2}+O\left(\frac{s^2}{v^4}\right).
\label{GBamp}
\end{equation}
Of course, the Goldstone bosons are not directly observable, since through
the Higgs mechanism 
they are ``eaten'' by the $W^\pm$ and $Z$ longitudinal components,
that we will denote, generically, by $W_L$.
Indeed, the so-called 
Equivalence Theorem \cite{et} (ET) relates the Goldstone bosons
 amplitudes with the corresponding 
longitudinal components of the electroweak
 gauge bosons for the MSM, as follows
\begin{equation}
A(W^a_LW^b_L\rightarrow W^c_LW^d_L)\simeq A(\omega^a
\omega^b \rightarrow \omega^c
\omega^d)+O\left(\frac{M_W}{\sqrt{s}}\right).
\label{WWamp}
\end{equation}
This result is a consequence of the Slavnov-Taylor
identities coming from the $SU(2)_L\times U(1)_y$ gauge symmetry.
The $O(M_W/\sqrt(s))$ corrections can be understood by noting
that the Goldstone bosons are massless in contrast with
the gauge bosons, whose mass is $O(100\, \hbox{GeV})$.
Note that the ET is a high energy limit, whereas the EChL is a low energy
limit. Indeed, for the EChL the formulation of the Equivalence Theorem   
is not so simple \cite{ETeff} 
but we will not discuss the details here since, later
we will unitarize the amplitudes of the effective lagrangian and in 
such case the above formilation is valid (the interested reader can
 find a complete account of this issue in \cite{libro} and \cite{aet}). 
At this point the following comments are in order:

\begin{itemize}
\item\
First we see that the low energy dynamics  of the Goldstone bosons
 is dictated by symmetry and the scale $v$ only. In this sense,
it is {\em universal}, i.e. independent
of the details of the SBS. The amplitudes obtained
from eq.(\ref{L2}) are
 called the Low Energy Theorems.

\item\ The amplitudes above grow with the energy. Thus, if we
assume that no other particles modify this behavior
at low energies, they give rise 
to strong interactions for the Goldstone bosons
as well as for the
 longitudinal components of the electroweak gauge bosons, according
 to the Equivalence Theorem.

\item However, the growth of this amplitudes is
 in conflict with unitarity around $O(1 \hbox{TeV})$ energies.  

\end{itemize}

In conclusion, provided there are no 
 other light modes in the SBS, we expect
strongly interacting $W_L's$.
 From unitarity constraints we also expect 
 new physics at the TeV scale, possibly in the 
form of resonances.

\section{Beyond the Low Energy Theorems}

In order to switch on the gauge fields in the low energy EChL we 
change the derivatives in eq.(\ref{L2}) into the appropriate covariant 
derivatives, $D_\mu$, containing the electroweak gauge fields.
That is
\begin{equation}
{\cal L}_2=\frac{v^2}{4}\tr D_{\mu}U (D^{\mu}U)^\dagger.
\end{equation}
In addition we can introduce the next to leading order (four derivative) 
terms \cite{dh} in the EChL
\begin{equation}
{\cal L}_4=L_1 \left(\tr D_\mu U D^\mu U^\dagger\right)^2+
L_2 \left(\tr D_\mu U D^\nu U^\dagger\right)^2 + ...
\label{L4}
\end{equation}
where we have only displayed those terms that
give rise to $O(s^2/v^4)$ contributions
to the  Goldstone boson
elastic scattering amplitude. 
These $O(s^2/v^4)$ terms depend on several 
$L_i$ constants, which parameterize
 our ignorance on the SBS. For special values we recover some
 particular models. For example, the MSM with a 1 TeV Higgs corresponds
 to $L_1=0.007$ and $L_2=-0.0022$ whereas the simplest TC model with three
 technicolors has $L_1=-0.001$ and $L_2=0.001$. In addition, under
 renormalization these parameters can absorb the divergences appearing
 in the one-loop contributions to the amplitudes coming from ${\cal L}_2$,
 which are $O(s^2/v^4)$. 
From precision test of
 the SM it is possible to set bounds on some of these parameters. 
However, these bounds are too weak for
$L_1$ and $L_2$, which are expected to lie in the $10^{-3}$ to $10^{-2}$
range. That precision could only be reached after a few years of
LHC running at full luminosity.

\section{Unitarization and dispersion Relations}
Customarily, the longitudinal gauge boson amplitudes are
given in a basis of states of definite angular momentum, $J$,
and the ``weak isospin'', $I$, associated to the SU(2) group.
These ``partial waves'', $t_{IJ}$ are also obtained as an expansion
of the form
\begin{equation} 
t_{IJ}(s)=t^{(2)}_{IJ}(s)+t^{(4)}_{IJ}(s)+O(s^3),
\label{expansion}
\end{equation}
where the superscript refers to the corresponding energy (momentum)
 power. As we have already seen, they grow with the energy and
violate unitarity around $1\,\hbox{TeV}$.
In this basis, the elastic unitarity constraint can be easily written 
for physical values of $s$; it reads
\begin{equation}
\hbox{Im}\, t_{IJ}(s) =\mid t_{IJ}(s)  \mid ^2\quad \Rightarrow\quad
\hbox{Im}\,\frac{1}{t_{IJ}(s)}=-1,
\label{tunit}
\end{equation}
which is basically the Optical Theorem.
Although the results obtained from 
the Chiral Lagrangian break
unitarity, they are nevertheless unitary in the perturbative sense
\begin{equation}
\hbox{Im} t_{IJ}^{(4)}(s) = \mid t_{IJ}^{(2)}(s)  \mid ^2
\quad \Rightarrow\quad
 \frac{\hbox{Im}\, t_{IJ}^{(4)}(s)}{\mid t_{IJ}^{(2)}(s)  \mid ^2} = -1.
\label{pertunit}
\end{equation}
It is however possible to obtain unitary amplitudes from
the effective Lagrangian. Note that from eq.(\ref{tunit}) we know
exactly the imaginary part of the inverse of
the amplitude.
As a consequence, any unitary amplitude will satisfy
\begin{equation}
\frac{1}{t_{IJ}(s)}=\hbox{Re}\,\frac{1}{t_{IJ}(s)}-i\quad \Rightarrow\quad
t_{IJ}(s)=\frac{1}{\hbox{Re}\,t^{-1}_{IJ}(s) - i}.
\end{equation}
That is, {\em we only have to approximate the real part of the
inverse of the amplitude} $t^{-1}_{IJ}$,
by means of eq.(\ref{expansion}). Formally: 
$\hbox{Re}\,t^{-1}_{IJ}= (t^{(2)}_{IJ})^{-1}[1-\hbox{Re} t_{IJ}^{(4)}/t^{(2)}_{IJ}+...\,]$. 
Finally, using eq.(\ref{expansion}) we can write
\begin{equation}
t_{IJ}(s)=\frac{t^{(2)}_{IJ}}{1-t_{IJ}^{(4)}/t^{(2)}_{IJ}}
\end{equation}
which is known as the Inverse Amplitude Method (IAM).
It can be derived alternatively, by writing a two subtracted  dispersion
relation for the inverse amplitude. Using some extra hypothesis
and approximations it is possible to solve the dispersion relation
for $t_{IJ}(s)$ to find the same result.

This partial wave is strictly unitary and has the proper analytical structure
 with the appropriate cuts. In addition it is able to reproduce
poles which can be interpreted
 as resonances generated dynamically. Note also that, by expanding
 this amplitude in power of $s$, 
{\em we recover the chiral low-energy expansion}.

\subsection{The Inverse Amplitude Method at work}
The IAM method has been successfully applied
in a completely different physical context: the low-energy hadron 
dynamics \cite{iam}. As it is well know, QCD is the proper theory
 to describe strong interactions, but it cannot be applied directly at
 low energies due to the breaking of standard perturbation theory.
However in the limit where the three lightest quarks are massless,
 the QCD Lagrangian possesses a global symmetry (chiral symmetry) which 
rotates right quarks or left quarks between themselves. The symmetry group 
is $SU(3)_L \times SU(3)_R$ and for different reasons it is known that 
it is spontaneously broken to the $SU(3)_{L+R}$ group. The corresponding
 Goldstone bosons are identified with the pseudoscalar mesons $\pi^0,
 \pi^{\pm}, K^0, \bar{K^0} $ and $\eta$ and their relative low physical 
masses compared with the typical hadronic scale of 1 GeV,
can be considered as a perturbation effect due to the very small, but
non-zero, quark masses. Note that the symmetry pattern is very close
to that of the SBS (it would be the same if we just considered
two quarks). 

As we did before we can gather
the $\omega^a$ mesons fields in an $SU(3)$ matrix as
$U(x)=\exp (i\omega^a \lambda^a/F)$, where $\lambda^a$ are the
 Gell-Man matrices and $F$ is basically the pion decay constant.
Once more we can
describe the low energy hadron dynamics in terms of a
chiral Lagrangian. This approach is known as Chiral Perturbation 
Theory (ChPT)\cite{chpt}.
At the lowest order this Lagrangian is given by:
\begin{equation}
{\cal L}_2=\frac{F^2}{4}\tr \,\partial_{\mu}U\partial^{\mu}U^\dagger.
\end{equation}
which reproduces the well know current algebra results in a very 
simple way. At the next order (four derivatives) one has additional terms
whose precise form is not relevant here, although some of them have the
same structure of those in eq.(\ref{L4}). 
As a matter of fact, the formalism that we have presented for the SBS is
inspired in the massless limit of SU(2) ChPT, although
rescaled from $F\simeq93\, \hbox{MeV}$ up to $v\simeq250\,\hbox{GeV}$.
The main difference is the existence of real data on meson physics, 
from which it is possible to determine the values of the ${\cal L}_4$
ChPT Lagrangian parameters, whereas they are undetermined for the SBS.

The amplitudes can now
be obtained as a truncated series in powers of
 the momentum $p^2$ over $4\pi F\simeq 1 \,\hbox{GeV}$. 
This formalism is only suitable at low-energies
up to about $500 \,\hbox{MeV}$.
We should not extrapolate them naively
to higher energies since
 they would severely violate unitarity and they would
not reproduce resonances. 

\begin{figure}[t]
\hspace*{2cm}
\hbox{\psfig{figure=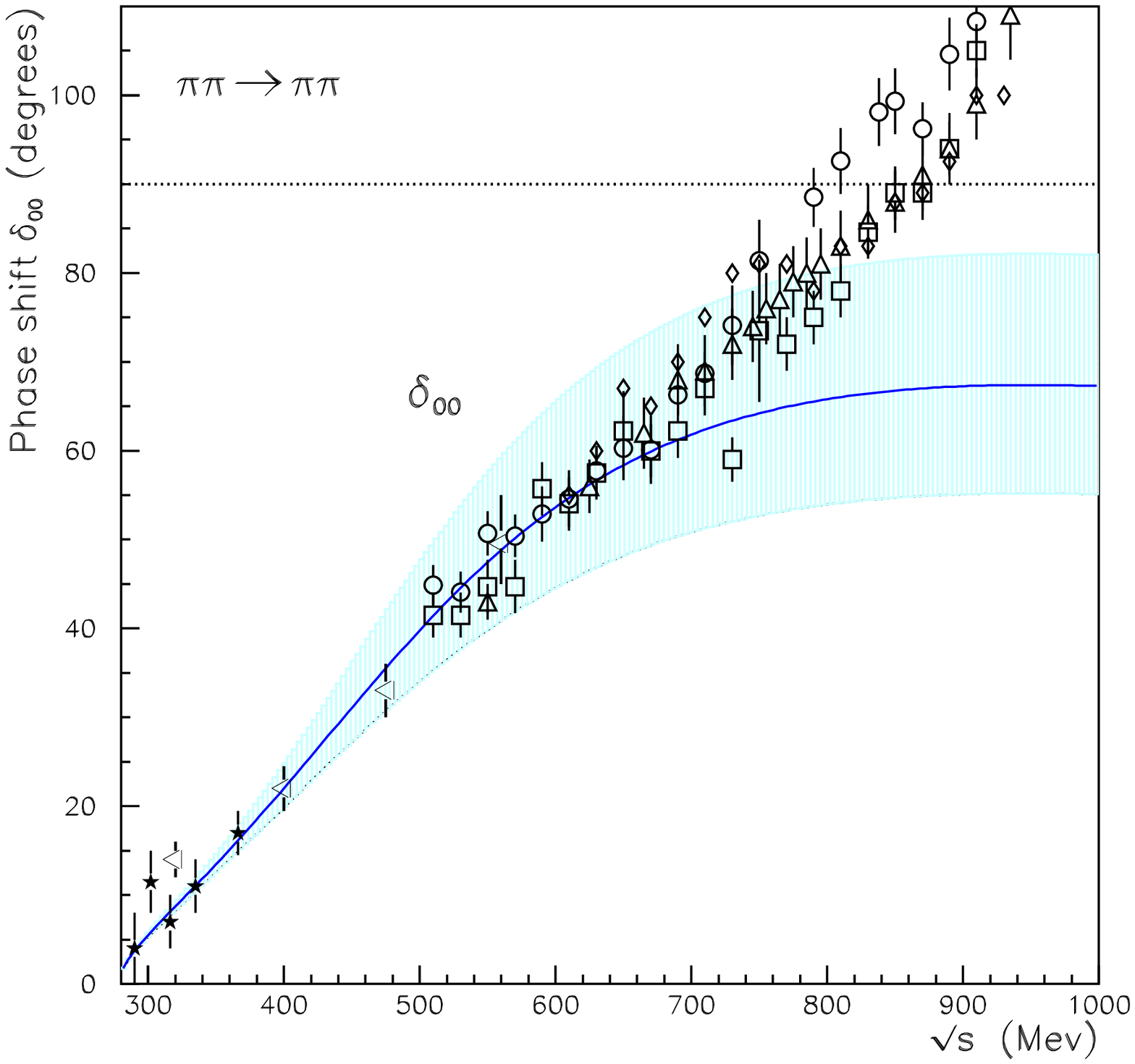,height=4cm}
\psfig{figure=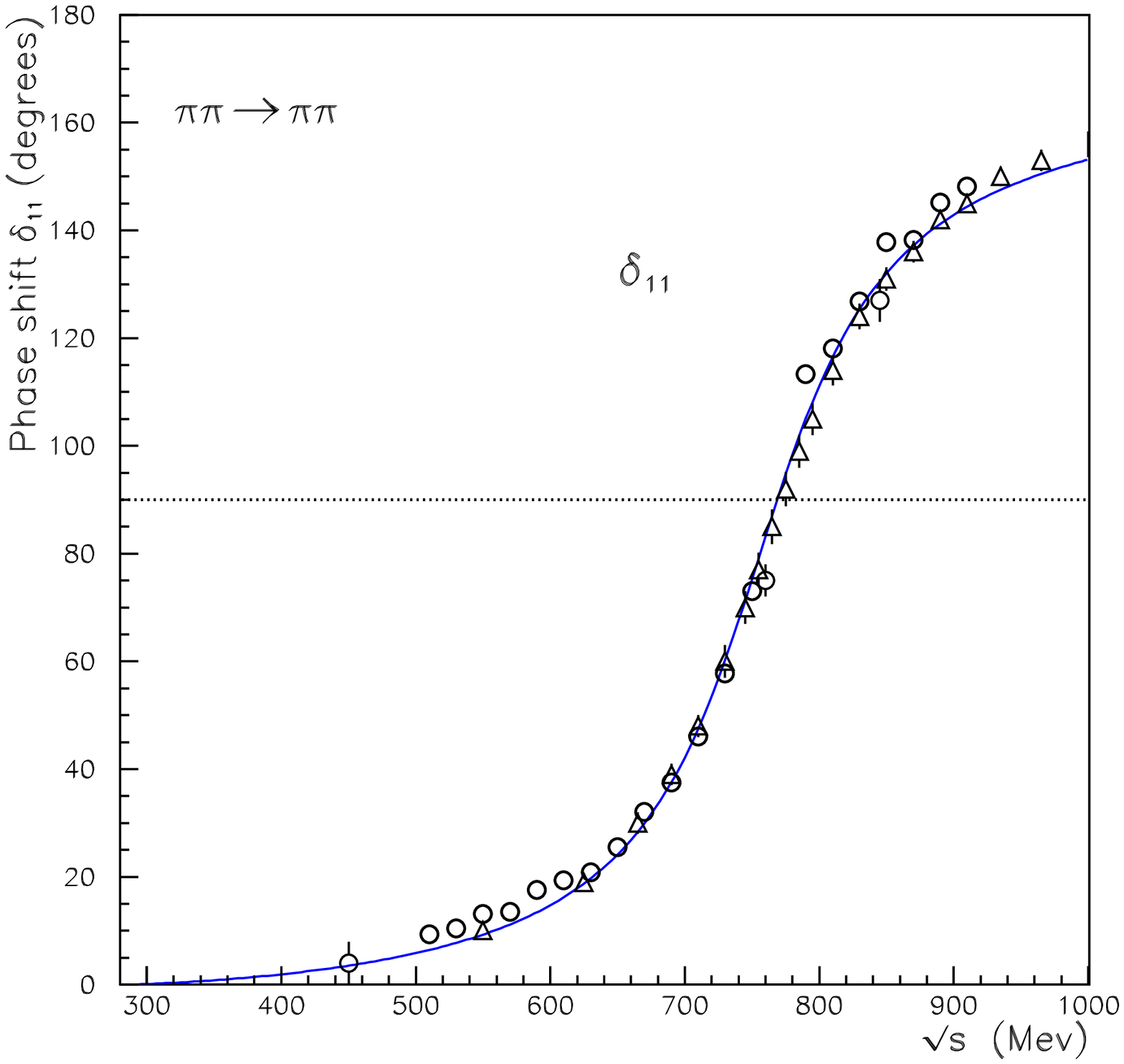,height=4cm}
\psfig{figure=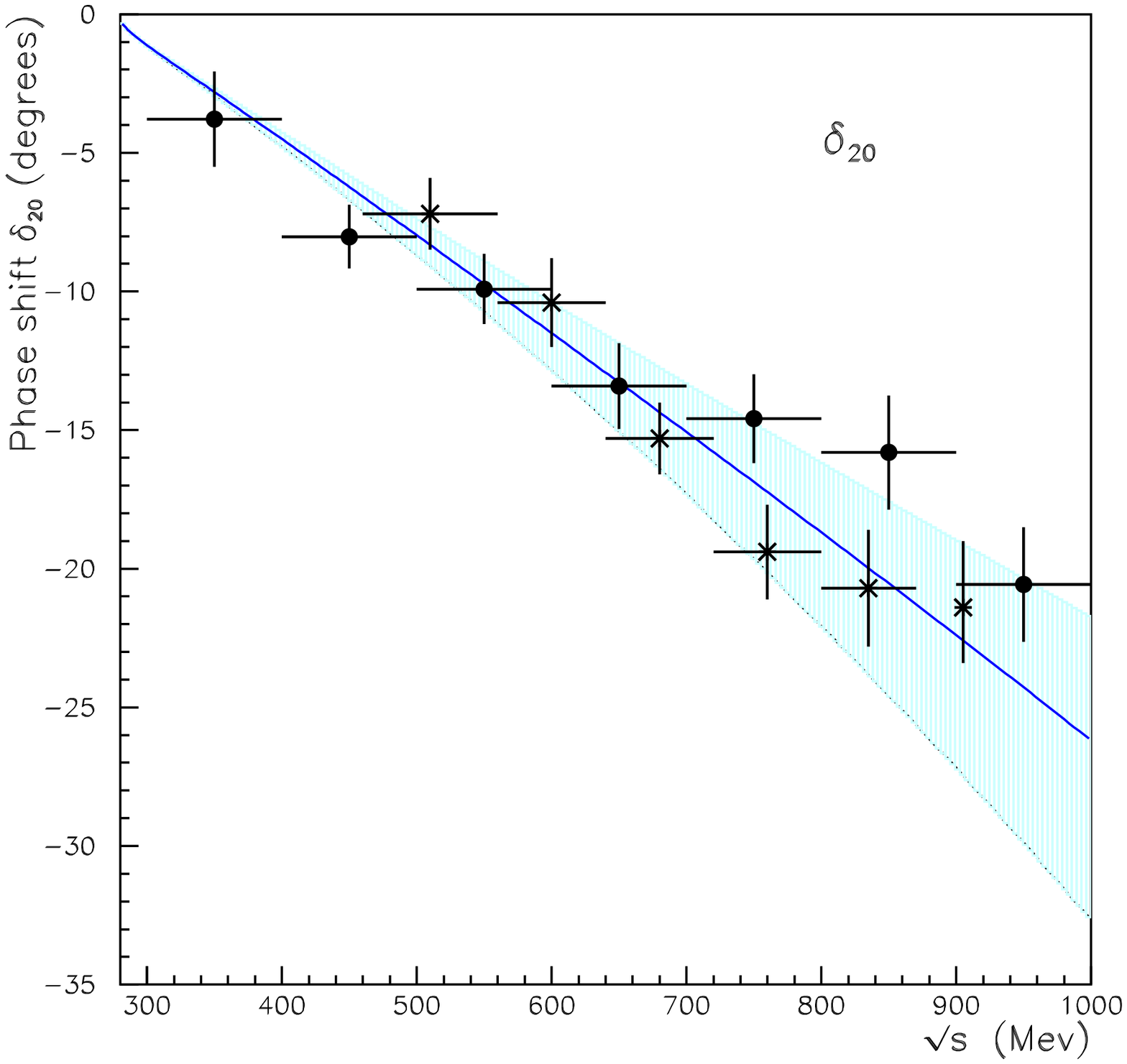,height=4cm}}

\hspace*{2cm}
\hbox{\psfig{figure=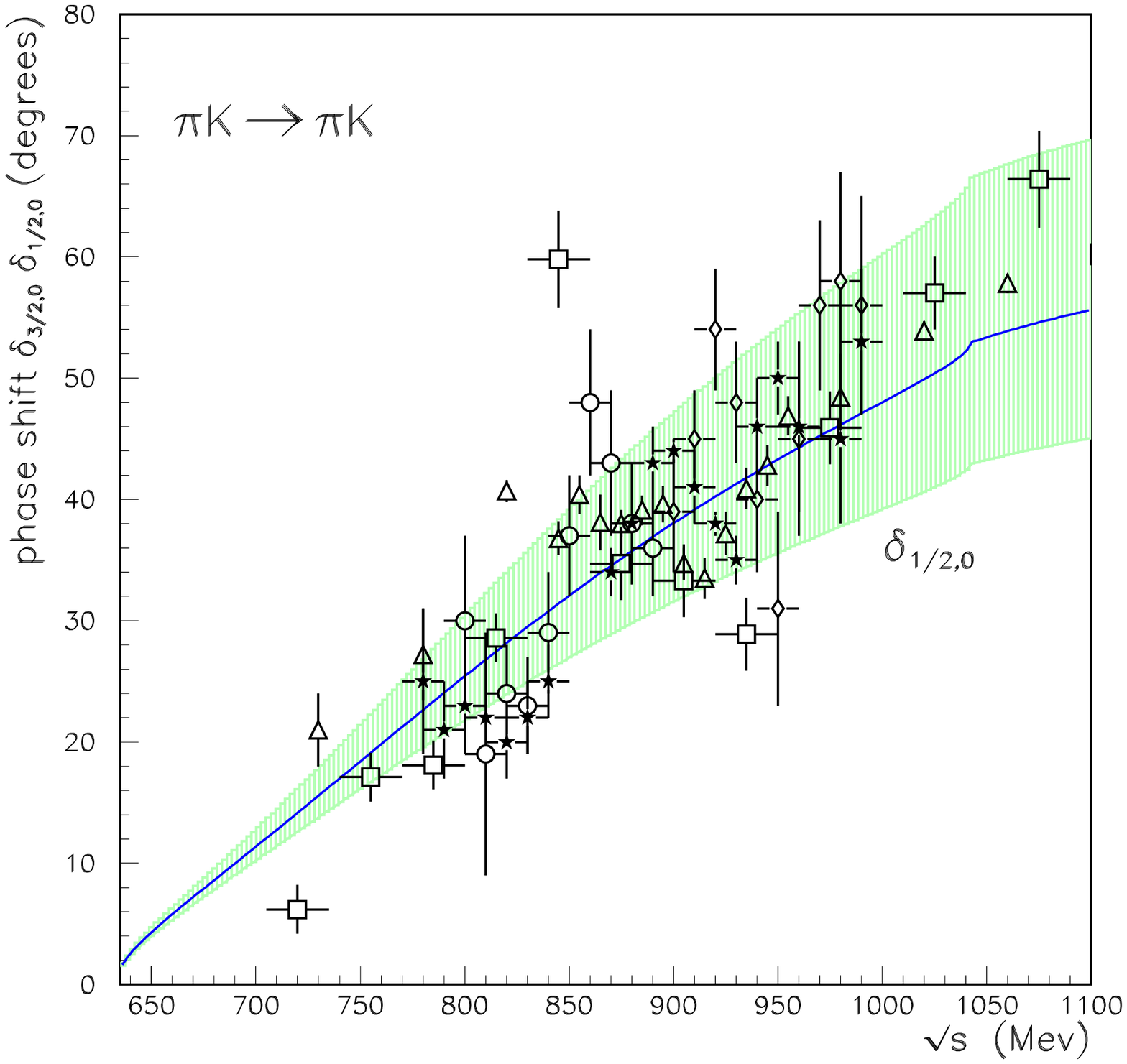,height=4cm}
\psfig{figure=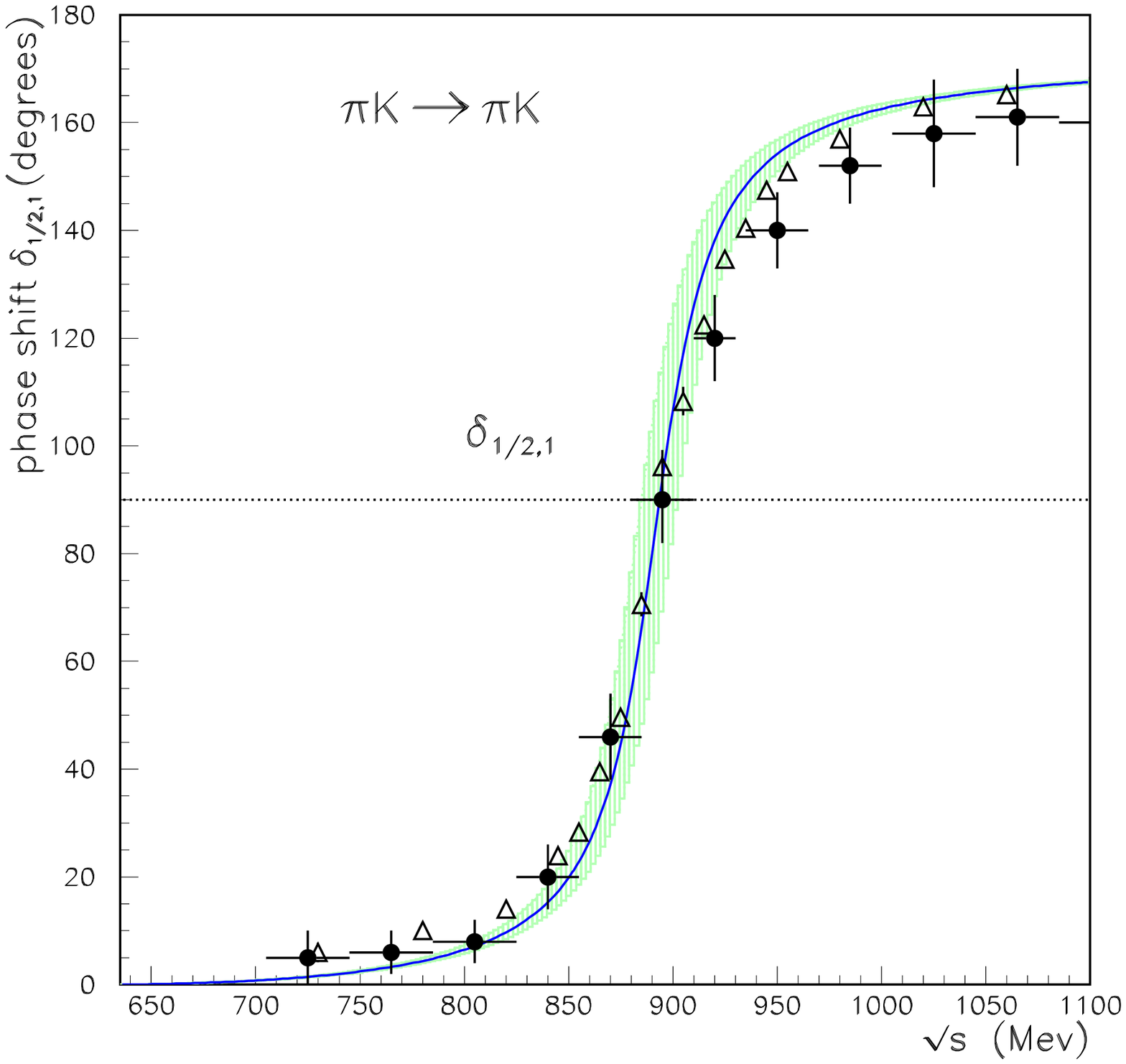,height=4cm}
\psfig{figure=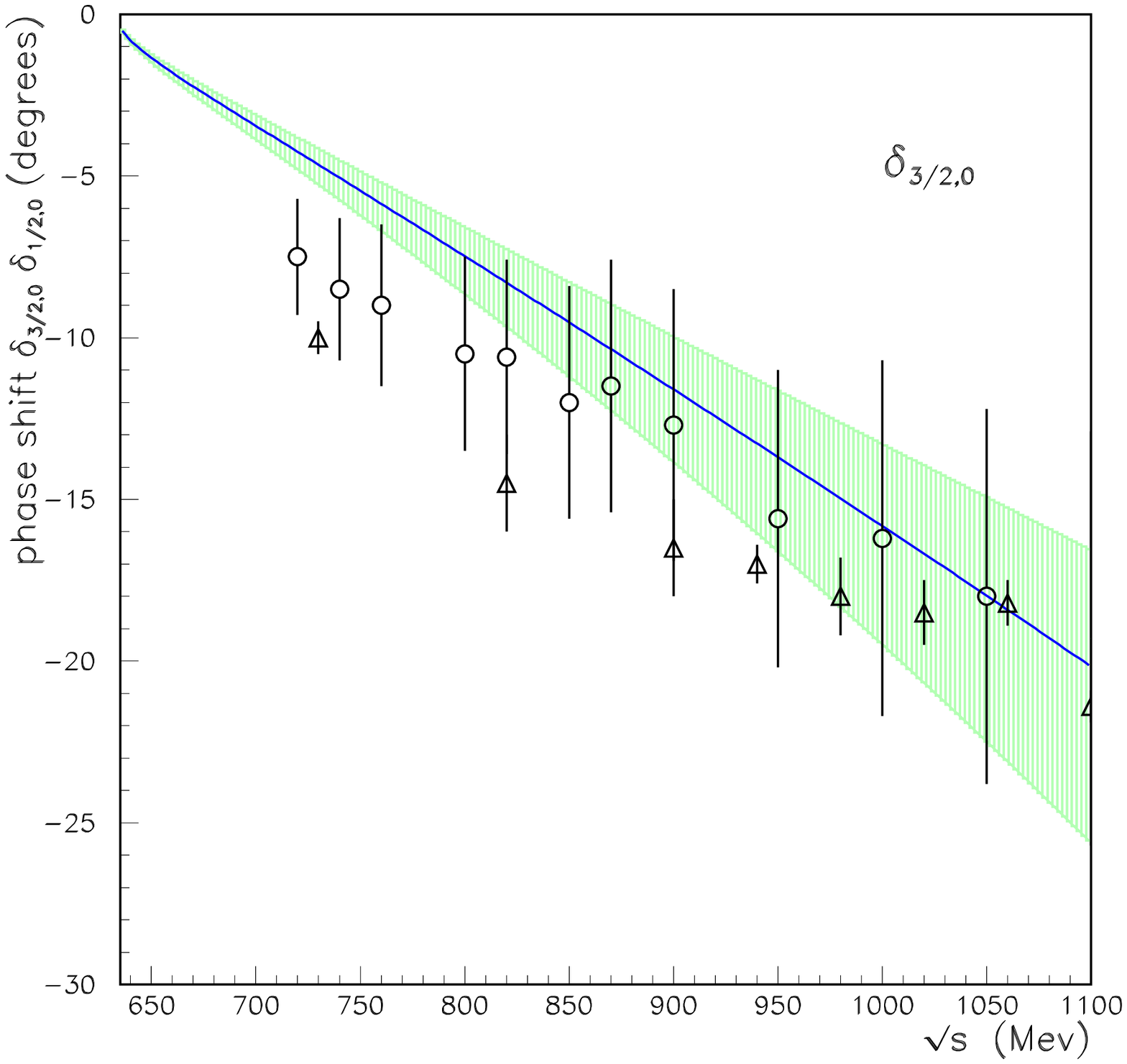,height=4cm}}
\caption{ IAM fit to the phase shifts for $\pi\pi\rightarrow\pi\pi$ and
$\pi K\rightarrow\pi K$. For data references see$^5$. }
\end{figure}
However, we can use the IAM to extend the applicability
of the effective Lagrangian approach. In Fig.1 we show an example
of the results obtained with the IAM when applied to 
$\pi\pi\rightarrow\pi\pi$ (analogous to the Goldstone 
boson $\omega\omega\rightarrow\omega\omega$) and $\pi K\rightarrow\pi K$
scattering. Note that now the data is reproduced up to approximately
1 GeV. In addition,
resonances like the $\sigma$, $\rho$ and $K^*$ are correctly
reproduced with an associated pole in
the second Riemann sheet.

Starting from the corresponding effective Lagrangians,
the IAM has also been applied very successfully to other processes
with coupled channels\cite{PRL} or even nucleons\cite{Oset}, 
reproducing correctly 
many other resonances. Thus we arrive to the conclusion that the IAM greatly
 improves the range of applicability of the effective Lagrangians
 and, moreover, it 
is able to reproduce resonances in the channels where they 
are present.

\section{Resonances in the SBS}

Let us then apply the IAM to the SBS. In this case we do not
have experimental information yet, and therefore we do not know
the values of the $O(p^4)$ parameters, which are model dependent.
Nevertheless, we expect them to lie between  $10^{-2}$  and $10^{-3}$
if the SBS is strongly interacting. 

For elastic $W_LW_L$ scattering only two $O(p^4)$ parameters
appear in the amplitudes, namely, $L_1$ and $L_2$. By changing their
values we can therefore reproduce the behavior of $W_LW_L$ scattering
in any strongly interacting model. In Fig.3 we show the phase shifts
$\delta_{IJ}$ that we expect in three 
different models. The first one corresponds to the SM with $1\, \hbox{TeV}$ 
Higgs and, consistently, we see a resonance in the scalar isoscalar
channel (a ``Higgs''). The second set of values mimics a simple TC model with
three technicolor and thus presents features very similar to ChPT (compare
with the $\pi\pi$ curves in Fig.1), mainly, a vector resonance (a ``techni-$\rho$).
The last model is chosen to show two behaviors that deserve further comments.
First, it could happen that a resonance becomes so broad that it
may be hard to identify as a resonance, in such case we say 
there is a ``saturation'' of unitarity. (The situation with the $\sigma$ 
particle in QCD is of this kind). Second, we have to remember that 
in the effective Lagrangian approach we only have a finite theory order
by order in energy, but it is not renormalizable in the strict sense.
Thus, the set of possible consistent fundamental theories is ``smaller''
 than that of effective theories. By that we mean that there could
be a choice of parameters which are {\em not} the low-energy limit of any
fundamental theory. That may seem obvious if we take an absurd
value for some parameters, like $L_1=10^6$. But it could also 
occur for values that look ``reasonable''. In such case, however, 
we would find  inconsistencies in the effective theory. That happens
indeed for the last model in Fig.3, which yields
a pole in the {\em first} Riemann sheet of the $I=2$ channel, 
which should not be present in
a renormalizable quantum field theory. That can be used as a criterium
to exclude a set of parameters. There are several arguments 
that support this interpretation\cite{yo}.

\begin{figure}
\hspace*{2cm}
\psfig{figure=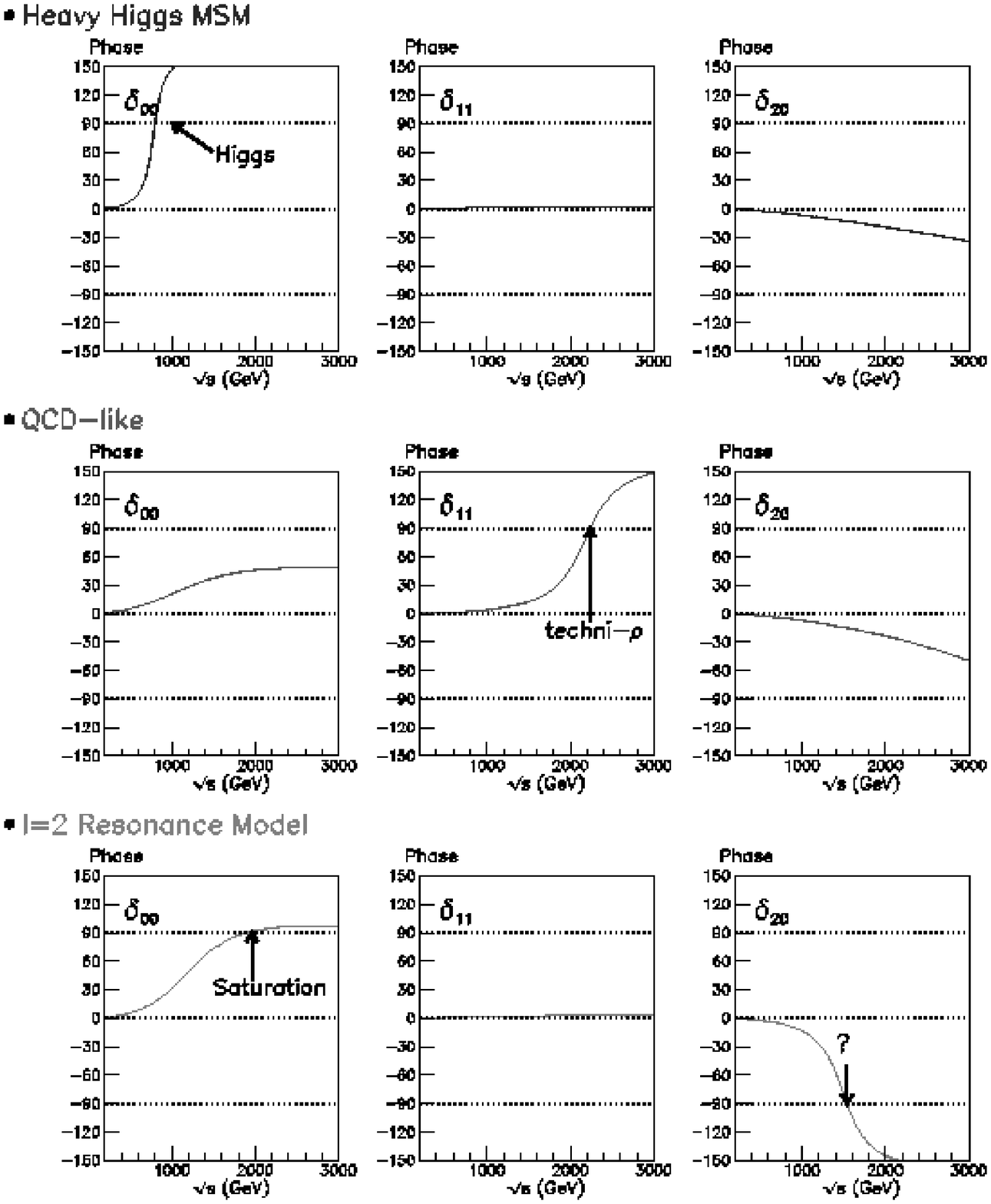,height=14cm}
\caption{Phase shifts expected for different choices of
electroweak chiral parameters.}
\end{figure}

We are now in conditions to study the general resonance spectrum of the 
strongly interacting SBS. We only have to vary the values of 
$L_1$ and $L_2$ in their expected ranges, and  identify what resonances 
appear below 3 TeV (an estimate of the LHC $W_LW_L$ scattering reach). 
In Fig.5 we presents the results of this approach\cite{yo},
which deserves some comments:

\begin{itemize}
\item\ 
The presence of an scalar resonance is represented by the areas
that contain an ``H'', whereas vector resonances are represented by 
a ``$\rho$''. Saturation effects are labeled by ``$S_I$'' for each
$I= 0, 1, 2$ channel. 

\item\
For illustrative purposes, we have signaled the pairs of
parameters that mimic some simple scenarios. The black triangles 
stand at the  position of a QCD-like model with 5 or 3 technicolors.
The black dots correspond to the SM with a Higgs whose tree level mass
is 800, 1000 or 1200 GeV.

\item\ 
Note that there are scenarios where we could find
two resonances in two different channels, or a resonance 
in one channel an a saturation behavior in another, or two 
saturation effects.

\item\
The black area is the part of parameter space which is excluded 
by the appearance of poles in the first Riemann sheet, it suggests that 
we cannot find {\em heavy} resonances in the $I=2$ channel 
(doubly charged Higgses). This difficulty has also been found when trying
to construct models with such particles: there is no model where they are
heavier than $\simeq 375\,\hbox{GeV}$ \cite{i2models}, 
a bound obtained from a renormalization
group analysis. In this lecture we are assuming that no such 
``light'' particles are present. In such case,
from the figure it seems that either nothing at all
or a ``saturation'' behavior is possible in that channel.

\item Finally, there is a small shaded region where no resonance or 
saturation effect would be clearly visible. In this region
it also seems  very hard to obtain a measurement of just
the chiral parameters and probably we would only get some bounds 
on their values \cite{MJ}. In such case, not even with the IAM we could get
any information of other more massive resonances that may lie ahead.

\end{itemize}

\begin{figure}[t]
\hspace*{3cm}
\psfig{figure=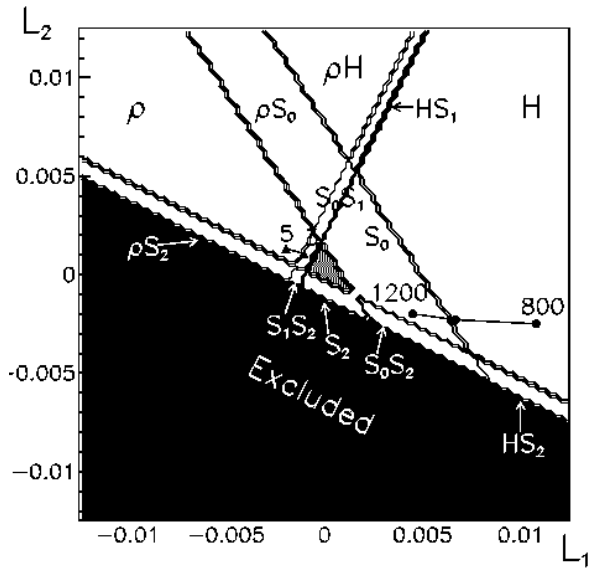,height=8cm}
\caption{Resonance spectrum of the strong SBS in the $L_1,L_2$
plane. The black area is excluded. On the white areas, we have
represented broad resonances or saturation effects
in the $I$ channel by $S_I$; Higgs-like narrow resonances by H and
$\rho$-like narrow resonances by $\rho$. In the grey area there is
no saturation of unitarity, nor resonances, below 3 TeV.
The black dots represent the MSM with $M_H=800,
1000, 1200 \mbox{GeV}$ and the triangles a QCD-like model with 3 or 5
technicolors. \label{fig:radish}}
\end{figure}

\section{Summary}

The main conclusions from the discussion below are the following:
\begin{itemize}

\item\
There is not any fundamental reason for the Higgs (Standard or
 Supersymmetric) to exist. We should therefore keep an open mind to 
alternative scenarios.

\item\
However unitarity requires new physics to appear before below
 1 TeV.

\item\
This new physics could be new particles in the best of the worlds
 or even a completely new and unexpected physics.

\item\
In the worst case we will have an enhancement of the $WW$ production.
 It will be difficult to observe at the LHC but not impossible. 
A lot of work should
 be done in this direction and chiral Lagrangians, supplemented with
the inverse amplitude method, can provide a model independent 
approach to new phenomena like the strong $W_LW_L$ scattering
and the resonances that may appear over the LHC energy range.

\item\ 
In any case we have to wait for the LHC with an open mind. Nature
 will tell us.

\end{itemize}

\section*{Acknowledgments}
This work has been partially supported by the Ministerio de Educaci\'on y
Ciencia (Spain)(CICYT AEN97-1693) and the U.S. Department of Energy 
under contract DE-AC03-76SF00515. J.R.P. thanks 
the Theory Group at SLAC for their kind hospitality.

\end{document}